\documentclass[twocolumn,times]{aastex63}
\usepackage{soul}
\usepackage{color, xcolor}
\usepackage{graphicx}
\usepackage{makecell}
\usepackage{amsmath}
\usepackage{amsthm}
\usepackage{CJKutf8}
\usepackage{booktabs}
\usepackage{lineno}
\usepackage{threeparttable}
\usepackage[squaren]{SIunits}
\usepackage{xspace}
\usepackage{comment}
\usepackage{xcolor}
\usepackage{multirow}
\usepackage{savesym}
\usepackage{amsmath}
\savesymbol{singlespace}
\savesymbol{doublespace}
\usepackage{txfonts}
\restoresymbol{TXF}{iint}
\usepackage{setspace}

\usepackage{array}

\setwatermarkfontsize{1.5in}

\shortauthors{Luo ET AL.}
\begin{document}
\begin{CJK}{UTF8}{gbsn}

\title{Pulsar and Magnetar Navigation with Fermi/GBM and GECAM}
\correspondingauthor{S. Xiao, S.L Xiong, S.N Zhang and Q.J Zhi}
\email{xiaoshuo@gznu.edu.cn, xiongsl@ihep.ac.cn,\\
zhangsn@ihep.ac.cn, qjzhi@gznu.edu.cn}
\author{Xi-Hong Luo}
\affil{Guizhou Provincial Key Laboratory of Radio Astronomy and Data Processing, Guizhou Normal University, Guiyang 550001, People’s Republic of China}
\affil{School of Physics and Electronic Science, Guizhou Normal University, Guiyang 550001, People’s Republic of China}

\author{Shuo Xiao*}
\affil{Guizhou Provincial Key Laboratory of Radio Astronomy and Data Processing, Guizhou Normal University, Guiyang 550001, People’s Republic of China}
\affil{School of Physics and Electronic Science, Guizhou Normal University, Guiyang 550001, People’s Republic of China}

\author{Shi-Jie Zheng}
\affil{Key Laboratory of Particle Astrophysics, Institute of High Energy Physics, Chinese Academy of Sciences, Beijing 100049, China}

\author{Ming-Yu Ge}
\affil{Key Laboratory of Particle Astrophysics, Institute of High Energy Physics, Chinese Academy of Sciences, Beijing 100049, China}

\author{You-Li Tuo}
\affil{Key Laboratory of Particle Astrophysics, Institute of High Energy Physics, Chinese Academy of Sciences, Beijing 100049, China}
\affil{Institut für Astronomie und Astrophysik, University of Tübingen, Sand 1, 72076 Tübingen, Germany}

\author{Shao-Lin Xiong*}
\affil{Key Laboratory of Particle Astrophysics, Institute of High Energy Physics, Chinese Academy of Sciences, Beijing 100049, China}

\author{Shuang-Nan Zhang*}
\affil{Key Laboratory of Particle Astrophysics, Institute of High Energy Physics, Chinese Academy of Sciences, Beijing 100049, China}
\affil{University of Chinese Academy of Sciences, Chinese Academy of Sciences, Beijing 100049, China}

\author{Fang-Jun Lu}
\affil{Key Laboratory of Particle Astrophysics, Institute of High Energy Physics, Chinese Academy of Sciences, Beijing 100049, China}

\author{Yue Huang}
\affil{Key Laboratory of Particle Astrophysics, Institute of High Energy Physics, Chinese Academy of Sciences, Beijing 100049, China}

\author{Cheng Yang}
\affil{Guizhou Provincial Key Laboratory of Radio Astronomy and Data Processing, Guizhou Normal University, Guiyang 550001, People’s Republic of China}
\affil{School of Physics and Electronic Science, Guizhou Normal University, Guiyang 550001, People’s Republic of China}

\author{Qi-Jun Zhi*}
\affil{Guizhou Provincial Key Laboratory of Radio Astronomy and Data Processing, Guizhou Normal University, Guiyang 550001, People’s Republic of China}
\affil{School of Physics and Electronic Science, Guizhou Normal University, Guiyang 550001, People’s Republic of China}

\author{Li-Ming Song}
\affil{Key Laboratory of Particle Astrophysics, Institute of High Energy Physics, Chinese Academy of Sciences, Beijing 100049, China}

\author{Wen-Xi Peng}
\affil{Key Laboratory of Particle Astrophysics, Institute of High Energy Physics, Chinese Academy of Sciences, Beijing 100049, China}

\author{Xiang-Yang Wen}
\affil{Key Laboratory of Particle Astrophysics, Institute of High Energy Physics, Chinese Academy of Sciences, Beijing 100049, China}

\author{Xin-Qiao Li}
\affil{Key Laboratory of Particle Astrophysics, Institute of High Energy Physics, Chinese Academy of Sciences, Beijing 100049, China}

\author{Zheng-Hua An}
\affil{Key Laboratory of Particle Astrophysics, Institute of High Energy Physics, Chinese Academy of Sciences, Beijing 100049, China}

\author{Jin Wang}
\affil{Key Laboratory of Particle Astrophysics, Institute of High Energy Physics, Chinese Academy of Sciences, Beijing 100049, China}

\author{Ping Wang}
\affil{Key Laboratory of Particle Astrophysics, Institute of High Energy Physics, Chinese Academy of Sciences, Beijing 100049, China}

\author{Ce Cai}
\affil{College of Physics, Hebei Normal University, 20 South Erhuan Road, Shijiazhuang, 050024, China}

\author{Cheng-Kui Li}
\affil{Key Laboratory of Particle Astrophysics, Institute of High Energy Physics, Chinese Academy of Sciences, Beijing 100049, China}

\author{Xiao-Bo Li}
\affil{Key Laboratory of Particle Astrophysics, Institute of High Energy Physics, Chinese Academy of Sciences, Beijing 100049, China}

\author{Fan Zhang}
\affil{Key Laboratory of Particle Astrophysics, Institute of High Energy Physics, Chinese Academy of Sciences, Beijing 100049, China}

\author{Ai-Jun Dong}
\affil{Guizhou Provincial Key Laboratory of Radio Astronomy and Data Processing, Guizhou Normal University, Guiyang 550001, People’s Republic of China}
\affil{School of Physics and Electronic Science, Guizhou Normal University, Guiyang 550001, People’s Republic of China}

\author{Wei Xie}
\affil{Guizhou Provincial Key Laboratory of Radio Astronomy and Data Processing, Guizhou Normal University, Guiyang 550001, People’s Republic of China}
\affil{School of Physics and Electronic Science, Guizhou Normal University, Guiyang 550001, People’s Republic of China}

\author{Jian-Chao Feng}
\affil{Guizhou Provincial Key Laboratory of Radio Astronomy and Data Processing, Guizhou Normal University, Guiyang 550001, People’s Republic of China}
\affil{School of Physics and Electronic Science, Guizhou Normal University, Guiyang 550001, People’s Republic of China}

\author{Qing-Bo Ma}
\affil{Guizhou Provincial Key Laboratory of Radio Astronomy and Data Processing, Guizhou Normal University, Guiyang 550001, People’s Republic of China}
\affil{School of Physics and Electronic Science, Guizhou Normal University, Guiyang 550001, People’s Republic of China}

\author{De-Hua Wang}
\affil{Guizhou Provincial Key Laboratory of Radio Astronomy and Data Processing, Guizhou Normal University, Guiyang 550001, People’s Republic of China}
\affil{School of Physics and Electronic Science, Guizhou Normal University, Guiyang 550001, People’s Republic of China}

\author{Lun-Hua Shang}
\affil{Guizhou Provincial Key Laboratory of Radio Astronomy and Data Processing, Guizhou Normal University, Guiyang 550001, People’s Republic of China}
\affil{School of Physics and Electronic Science, Guizhou Normal University, Guiyang 550001, People’s Republic of China}

\author{Xin Xu}
\affil{Guizhou Provincial Key Laboratory of Radio Astronomy and Data Processing, Guizhou Normal University, Guiyang 550001, People’s Republic of China}
\affil{School of Physics and Electronic Science, Guizhou Normal University, Guiyang 550001, People’s Republic of China}

\author{Meng-Xuan Zhang}
\affil{School of Physics and Electronic Science, Guizhou Normal University, Guiyang 550001, People’s Republic of China}

\author{Zi-Ping Dong}
\affil{School of Physics and Electronic Science, Guizhou Normal University, Guiyang 550001, People’s Republic of China}

\author{Shi-Jun Dang}
\affil{Guizhou Provincial Key Laboratory of Radio Astronomy and Data Processing, Guizhou Normal University, Guiyang 550001, People’s Republic of China}
\affil{School of Physics and Electronic Science, Guizhou Normal University, Guiyang 550001, People’s Republic of China}

\begin{abstract}
The determination of the absolute and relative position of a spacecraft is critical for its operation, observations, data analysis, scientific studies, as well as deep space exploration in general. A spacecraft that can determine its own absolute position autonomously may perform more than that must rely on transmission solutions. In this work, we report an absolute navigation accuracy of $\sim$ 20 km using 16-day Crab pulsar data observed with $Fermi$ Gamma ray Burst Monitor (GBM). In addition, we propose a new method with the inverse process of the triangulation for joint navigation using repeated bursts like that from the magnetar SGR J1935+2154 observed by the Gravitational wave high-energy Electromagnetic Counterpart All-sky Monitor (GECAM) and GBM.

\end{abstract}

\keywords{navigation - pulsars - magnetars}

\section{Introduction}
X-ray pulsar-based navigation (XPNAV) has been considered important and been actively developed for both orbit estimation of Earth satellites and deep space navigation for spacecraft, since it was proposed by Chester and Butman in 1981 \citep{chester1981navigation}. 
Zheng et al. \citep{2017SSPMA..47i9505Z} proposed the `Significance Enhancement of Pulse-profile with Orbit-dynamics' (SEPO) method and successful applied it for the first time to a pulsar navigation experiment, with the data from the observations of the Crab pulsar obtained by the POLAR experiment \citep{produit2018design} onboard China’s Tiangong-2 spacelab. 
Subsequently, the same method was applied to the first Chinese space X-ray astronomical telescope called the $Insight$-Hard X-ray Modulation Telescope ($Insight$-HXMT) \citep{zhang2020overview} to determine its orbit position with accuracy of 10 km (3$\sigma$) \citep{zheng2019orbit}, and a navigation experiment has been conducted successfully with the Gravitational wave high-energy Electromagnetic Counterpart All-sky Monitor (GECAM) \citep{gecam_dh} using this method. Although in the method only the deviation of a single orbital element is considered \citep{fang2021analysis}, it is nevertheless a convenient way to navigate and avoid the effects of combining different pulsars. On the other hand, navigation accuracy within 5 km (1$\sigma$) has been achieved with the Station Explorer for X-ray Timing And Navigation Technology (SEXTANT) \citep{witze2018nasa} implemented to the Neutron star Interior Composition Explorer (NICER) by observing five millisecond pulsars (\citealp{winternitz2015x}).

Similar to POLAR's success in the field of gamma-ray
burst (GRB) polarization measurements \citep{zhang2019detailed}, the $Fermi$ Gamma ray Burst Monitor (GBM) \citep{meegan2009fermi} is a greatly successful high-energy transient detector that has made important contributions since its launch in 2008, such as the first observation of the electromagnetic signal of a binary neutron star merger GRB 170817A \citep{goldstein2017ordinary}. 
GBM is composed of 12 sodium iodide (NaI) scintillators and 2 bismuth germanate (BGO) scintillators to study Gamma-Ray Bursts (GRBs) in 8 keV to 40 MeV. The NaI detectors measure the low-energy photons (8 keV to 1 MeV), and the crystal disks have a diameter of 12.7 cm (5 inches) and a thickness of 1.27 cm (0.5 inches). The absolute timing of the GBM clock has accuracy better than 10 $\rm \micro\second$, and the time resolution is 2 $\rm \micro\second$ (\citealp{meegan2009fermi}; \citealp{paciesas2012fermi}). However, to date, GBM has not been used for pulsar navigation experiments.

On the other hand, XPNAV has the disadvantage of requiring a long observation time (e.g., ten days) (\citealp{winternitz2015x}; \citealp{zheng2019orbit}) and a time-consuming computational effort (e.g., the arrival time of each photon must be corrected to that arriving at the solar system barycenter) \citep{2017SSPMA..47i9505Z}, as well as modeling the timing noise of pulsars (e.g., the red noise, spin evolution and glitches) (\citealp{lyne199323}; \citealp{scott2003characterization}; \citealp{hobbs2010analysis}; \citealp{lyne201545}). 
Fortunately, there are many sources that have repeated bursts in the universe, such as Soft Gamma-ray Repeaters (SGRs), a kind of gamma-ray transient sources that exhibit explosive activity from magnetars. For example, SGR J1935+2154 is one of the most active magnetars since its discovery in 2014 \citep{israel2016discovery} (e.g. hundreds of bursts were observed within April 27, 2020 \citep{li2021hxmt,zou2021periodicity,xie2022revisit}). In addition, the bursts from SGR J1935+2154 are very easy to identify, e.g. a typical duration of $\sim$ 0.1 seconds with energies mostly in the order of tens of keV.
Due to their high brightness, sharp light curves and tiny spectral lags between different energy bands of these bursts, they were well used to verify the time delay location and calibration of satellite time systems (\citealp{xiao2022energetic}; \citealp{xiao2022ground}). 

GECAM was launched on Dec 10, 2020, with the highest time resolution (0.1 $\rm \micro\second$) among all GRB detectors ever flown and high absolute time accuracy (about 6.04 $\rm \micro\second$) \citep{xiao2022ground}. The payload of each GECAM satellite consists of 25 Gamma-Ray Detectors (GRDs) and 8 Charged Particle Detectors (CPDs). GRDs are used to detect X/gamma-rays, a large-volume Lanthanum (III) bromide (LaBr$_3$) with a diameter of 76 mm and a thickness of 15 mm read out by a Silicon photomultipliers (SiPM) array are utilized in the GRD design (\citealp{2018JInst..13P8014L}; \citealp{zhang2019energy}; \citealp{chen2021design}; \citealp{2021arXiv210900235Z}). In addition, the energy response of GECAM is similar to that of GBM \citep{xiao2022energetic}, making it well suited for joint navigation. Therefore, in this work, we propose to utilize the repeated bursts like that from SGR J1935+2154 observed by GECAM and GBM for multi-satellite navigation.

In this article, we first perform the pulsar navigation for Fermi/GBM in Section 2. In Section 3, we then present the multi-satellite navigation method based on SGR J1935+2154 and the demonstration of the orbital accuracy of the joint GBM and GECAM navigation. A summary of this study is given in Section 4.

\begin{figure*}
    \centering
    \includegraphics[scale=0.8]{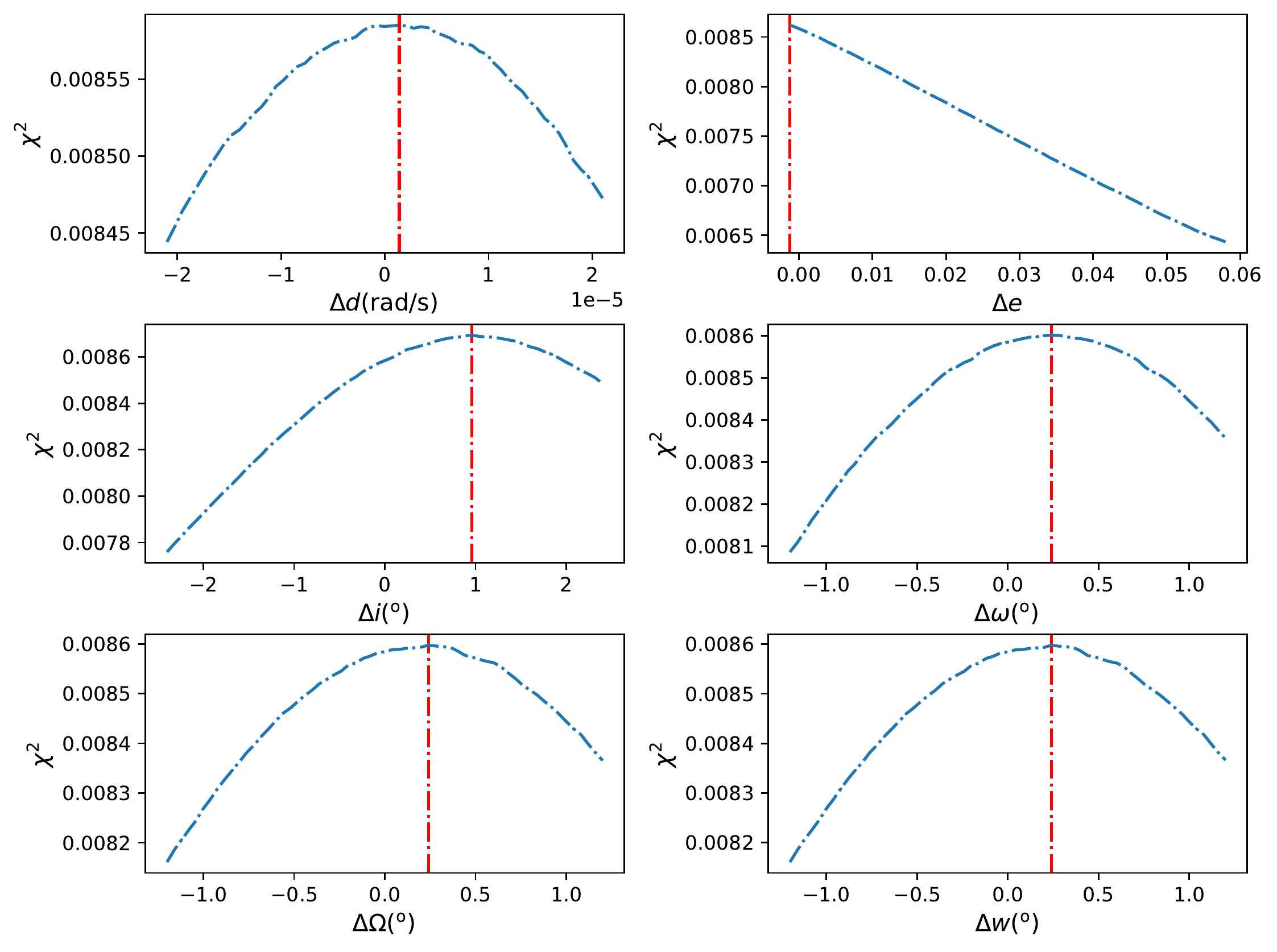}
    \caption{The ‘goodness’ of the pulse profile versus the variation of the six orbital elements. The Fermi/GBM 16-day observation for Crab from September 11 to 26, 2021 are used. The red dashed line in each panel represents the best value obtained based on pulsar navigation.}
    \label{kafang}
\end{figure*}

\begin{table*}[htbp]
\centering
\caption{\centering Six orbital elements of Fermi obtained by pulsar navigation. The deviations and errors are from Fig.~\ref{kafang} and Fig.~\ref{mcmc}, respectively.}\label{table1}
\begin{tabular}{c  c c c c}

\hline
Orbital Element Deviation     &Deviation &Error (1$\sigma$)  &Distance Deviation (km)  &Distance Error (km)
\\
\hline
$\Delta d${\rm (rad/s)}($10^{-3}$)     &0.0014           &0.00627        &7.412          &35.639
\\
$\Delta e$                &-0.00122       &(-0.00000 , +0.00022 )    &8.673      &(0.000 , +7.706 )                      \\
$\Delta i$${\rm (^o)}$          &0.96     &0.336      &32.841     &11.819                       \\
$\Delta \omega$${\rm (^o)}$     &0.24     &0.216      &29.328     &26.461                       \\
$\Delta \Omega$${\rm (^o)}$     &0.24     &0.225      &29.750     &28.270                       \\
$\Delta w$${\rm (^o)}$          &0.24     &0.225      &29.817     &27.984                       \\
\hline
\end{tabular}    
\end{table*}

\begin{figure*}
\centering
\includegraphics[scale=0.8]{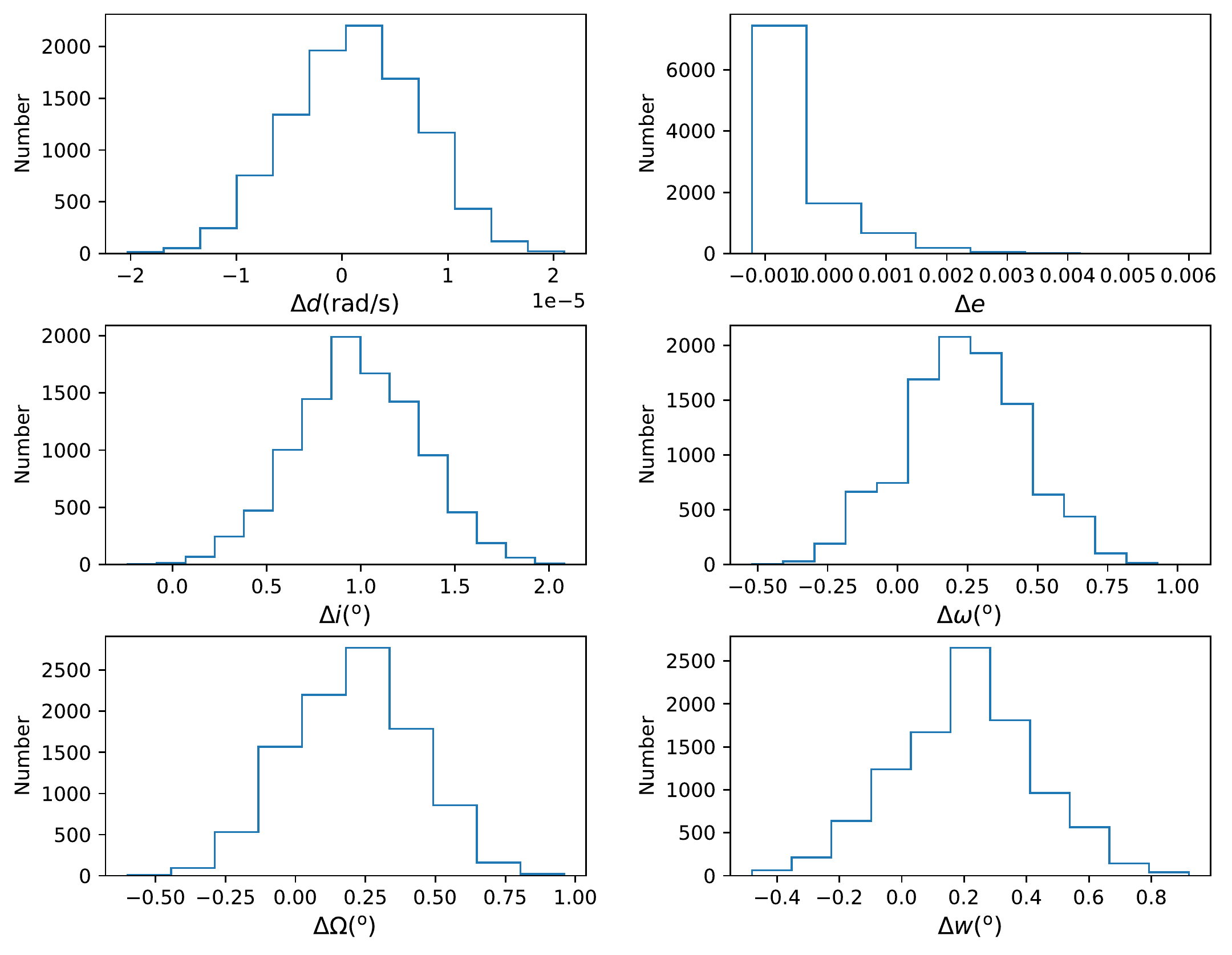}

\caption{Distribution of the six orbital elements obtained by the Monte Carlo method, from which the errors of these elements can be estimated.}\label{mcmc}
\end{figure*}

\begin{figure*}[t]
    \centering
    \includegraphics[scale=0.5]{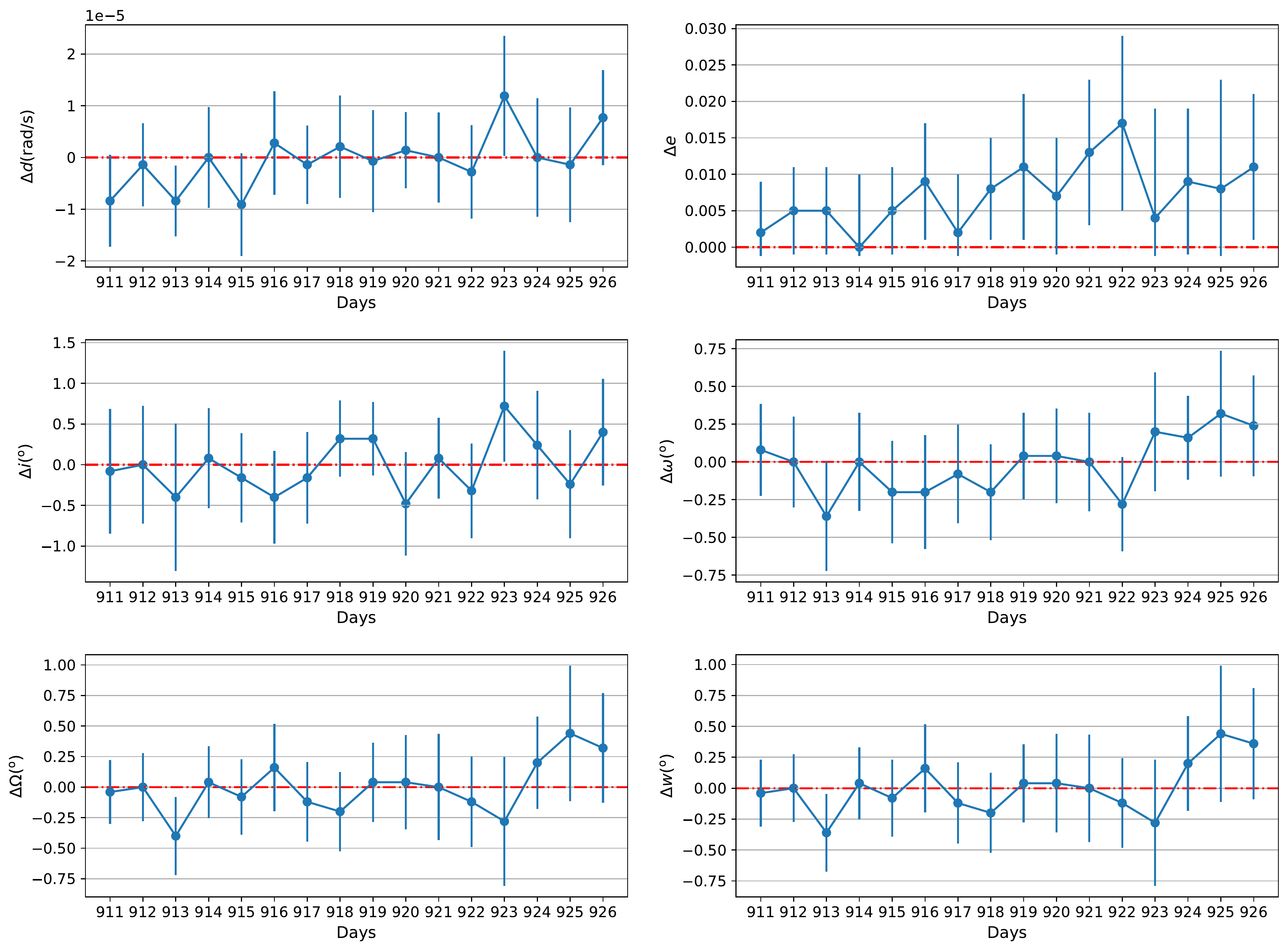}
    \caption{The six orbital elements obtained by daily Crab observations from September 11 to 26, 2021.}
    \label{each day}
\end{figure*}

\begin{figure*}[http]
\centering
\begin{minipage}[t]{0.99\textwidth}
\centering
\includegraphics[width=\columnwidth]{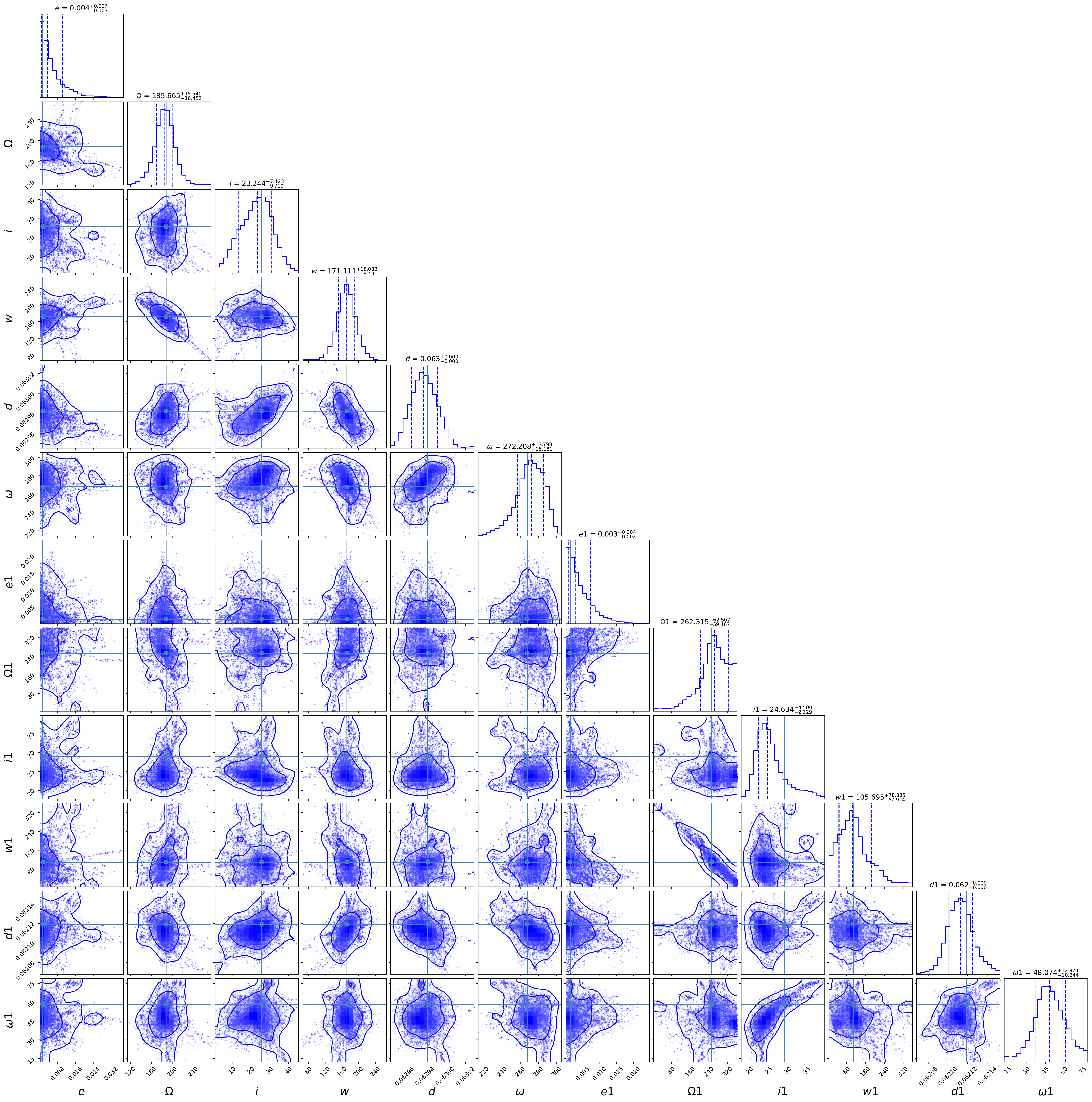}
\end{minipage}
    \caption{The magnetar navigation results obtained by MCMC. $e$, $\Omega$, $i$, $w$, $d$ and $\omega$ are the orbital elements eccentricity, argument of perigee, inclination angle, mean anomaly, mean motion and RAAN of Fermi/GBM, respectively. $e_1$, $\Omega_1$, $i_1$, $w_1$, $d_1$ and $\Omega_1$ are the orbital elements of GECAM-B, respectively. Detailed results are listed in Table \ref{table2}.}\label{mag_dh}
\end{figure*}

\section{Crab Pulsar navigation for Fermi/GBM}
\subsection{Methodology and Data treatment}
The Crab pulsar (PSR B0531+21, RA = $\rm 05^{h}34^{m}31.972^{s}$, Dec = $\rm 22 ^{\circ} 0'52".07$) \citep{lyne199323} has been widely used for pulsar navigation \citep{10.1117/12.156637} due to its relatively stable period and high flux. To achieve pulsar navigation, the procedure used
here was proposed by Zheng et al. \citep{2017SSPMA..47i9505Z, zheng2019orbit}. 
First, we can calculate the orbit of Fermi/GBM at any time using the $\rm package$ $sgp4$ \footnote{https://github.com/brandon-rhodes/python-sgp4}, according to the six orbital elements in the orbital dynamics model, which are Mean Motion ($d$), Eccentricity ($e$), Inclination Angle ($i$), RAAN (Right Ascension of the Ascending) ($\omega$), Argument of Perigee ($\Omega$), and Mean Anomaly ($w$).
However, the deviations between the calculated and actual orbits will become larger as the time difference increases. As shown in Fig. A1, when the time difference is 16 days, the bias in the predicted orbit is about 5 km. 
Therefore, in this work, we use the GBM 16-day observation for Crab (i.e. from September 11 to 26, 2021) to infer the six orbital elements at the moment of $T_0$ (i.e. 2021-09-11T13:31:13.426); the true values are $d=0.0629827\ {\rm (rad/s)}$, $e=0.0012175$, $i=25.5821\ {\rm (^o)}$, $\omega=267.808\ {\rm (^o)}$, $\Omega=188.04\ {\rm (^o)}$, $w=171.995\ {\rm (^o)}$. 
To improve the statistics, only those detectors on board GBM to which the incident angle of Crab pulsar is less than 70 degrees are used and those time intervals when the Crab pulsar is blocked by the Earth are ignored in this analysis, and only the events in 8-900 keV energy range of the NaI detectors are selected.

When we generate an orbit based on a certain set of six orbital elements, the arrival time of events observed by GBM are corrected to the solar system barycenter (DE200) based on the orbit. The phase $\Psi_{t}$ of each detected event is calculated with the following equation:

\begin{equation}
\begin{split}\label{equ:profile}
\Psi_{t}=\Psi_{0}+f_0(t-t_0)+\frac{1}{2}(t-t_0)^2 \times f_1+\frac{1}{6}(t-t_0)^3 \times f_2,
\end{split}
\end{equation}
where $f_0$, $f_1$ , $f_2$ are the radio ephemerides valid for the moment $t=t_0$ \footnote{http://www.jb.man.ac.uk/~pulsar/crab.html} \citep{lyne199323}. In this work, we make use of the $\rm package$ $tat-pulsar$ \footnote{https://github.com/tuoyl/tat-pulsar} \citep{tuo2022orbit} to obtain the pulse profile of the Crab pulsar (see Fig.~\ref{profiles_six}).

In order to infer the six orbital elements at the moment $t=t_0$, we generate a group of six orbital elements to produce different orbits. Then we use the following equation to evaluate the ‘goodness’ of the pulse profile based on these orbits, 
\begin{equation}
\begin{split}\label{equ:profile}
\chi^2 = \sum_{i}\frac{(p(\Psi_i)-\overline{p})^2}{\overline{p}^2},
\end{split}
\end{equation}
where $p(\Psi_i$) is the counts at the phase $\Psi_i$, $\overline{p}$ is the average of $p(\Psi$). It is worth noting that the denominator in the above equation is $\overline{p}^2$ to make the dimension consistent, which is different from the denominator $\overline{p}$ in \citealp{zheng2019orbit}.

As shown in Fig.~\ref{kafang}, we get the best values of the six orbital elements at the maximum value of $\chi^2$. To estimate the uncertainties, we make a Monte Carlo simulation of the observed counts of Crab profiles based on Poisson probability distribution, which is,
\begin{equation}
\begin{split}\label{equ:error}
\rm prob (n|\mu)=\frac{\mu^{n}}{n!}e^{-\mu},
\end{split}
\end{equation}
where $\mu$ is the observed counts, and $n$ is the counts obtained by sampling. The simulation is repeated 10000 times, and for each set of simulated profiles, we calculate the best set of the six orbital elements, respectively. 
The distributions of the differences of the simulated elements $\Delta d$, $\Delta e$, $\Delta i$ ,$\Delta \omega$, $\Delta \Omega$, and $\Delta  w$ approximates Gaussian (see Fig.~\ref{mcmc}), so that the standard deviation can be taken as the uncertainty. 
However, the distribution of the difference of the element $\Delta e$ is not Gaussian due to the fact that $e$ cannot be negative and $e$ for the Fermi satellite is close to 0; in this case, we take the range between the 0th and the 68th ranked values as its $1 \sigma$ confidence region.

\subsection{Absolute position of Fermi/GBM based on pulsar navigation}

Based on the Fermi/GBM 16-day observations of Crab pulsar, we obtain the pulse profiles at different sets of the six orbital elements and then the best values corresponding to the highest $\chi^2$ obtained, and the errors are obtained by the Monte Carlo sampling. 
The results are listed in Table \ref{table1}. Therefore, the orbital precision corresponding to the six orbital elements $d$, $e$, $i$, $\omega$, $\Omega$, $w$ are about 35.6, 7.8, 11.8, 26.5, 28.3 and 28.0 km, respectively; they are consistent with the true values within the 3$\sigma$ error range.

We also investigate the six orbital elements obtained from the daily data. As shown in Fig.~\ref{each day}, these results are consistent but the errors tend to increase as time increases (e.g. $\Delta w$ and $\Delta \Omega$ in Fig.~\ref{each day}), which is due to the fact that the predicted orbits calculated based on the orbital dynamics model will deviate from the true orbits more and more as time goes on. It is worth noting that we have used the data from $T_0$ to $T_0$+16 days to infer the six orbital elements at the moment $T_0$ instead of $T_0$-8 to $T_0$+8 days, so this will lead to larger elements errors.

\section{Magnetar navigation for Fermi/GBM and GECAM}
This method is in fact the inverse process of triangulation \citep{hurley2011interplanetary,pal2013interplanetary,2021ApJ...920...43X} for a transient source, that is, use these bursts to obtain the absolute and relative position of Fermi and GECAM-B by calculating the time delays between the burst arrival times to them. When a burst signal arrives at two spacecraft with a time delay $\Delta t$, it should satisfy the following equation,
\begin{equation}
\begin{split}\label{equ:Tri}
\rm cos \alpha=\frac{c\times \Delta t}{D} , 
\end{split}
\end{equation}
where $c$ is the speed of light, $D$ and $\alpha$ are respectively the distance between the two spacecraft and the incident angle with respect to the vector joining the two spacecraft, which contain the orbital information we need to navigate.
When the uncertainty in time delay $\sigma(\Delta t) \ll \rm D/c$, the uncertainty in $\alpha$ is
\begin{equation}
\begin{split}\label{equ:Trierr}
\sigma(\alpha)=\frac{c\times \sigma(\Delta t)}{D\times \rm sin \alpha}. 
\end{split}
\end{equation}
Therefore,  based on multiple bursts, we are able to obtain the orbital information and its accuracy from the time delays between the burst arrival times to the two satellites at different positions.

We utilize the Li-CCF method (i.e. Modified Cross Correlation Function (MCCF) method, \citealp{li2004timescale}) to calculate the time delay between the light curves observed by GECAM-B and GBM; the Li-CCF method can make the best of the temporal information in the high-resolution data and yield more accurate results \citep{li2004timescale,2021ApJ...920...43X}.

Table \ref{sgr_1935} shows the selected 26 bursts from SGR J1935+2154, as well as the calculated time delays and the true time delays observed by GBM and GECAM.
The probability distribution of the calculated time delay for the i-th burst in the sample is 
\begin{equation}\label{p_lag}
\begin{aligned}
p (i; \Delta t_{\rm mod})=\frac{1}{\sqrt{2\pi}\sigma_{\Delta t_{\rm obs}}} \exp\left[-\,\frac{(\Delta t_{\rm obs}-\Delta t_{\rm mod})^{2}}{2\left(\sigma_{\Delta t_{\rm obs}}\right)^{2}}\right],
      \end{aligned}
\end{equation}
where $\Delta t_{\rm obs}$ and $\sigma_{\Delta t_{\rm obs}}$ are the observed time delay and its error, respectively. $\Delta t_{\rm mod}$ is the expected value according to the model (i.e. the predicted delay based on orbital parameters).

Once any set of orbital elements is given, the predicted delay between the two satellites can be calculated based on the position of the satellite and the source. The orbital elements are calculated by the 
maximum likelihood estimation method, which is 
\begin{equation}\label{likelihood0}
\begin{aligned}
{\mathcal{L}(\theta) =\prod_{i} p (i; \Delta t_{\rm mod}(\theta))},
      \end{aligned}
\end{equation}
where $p$ is the probability distribution of the calculated time delay (Equation \ref{p_lag}), $\Delta t_{\rm mod}$ is the predicted delay, and $\theta$ is the orbital elements.

Fig.~\ref{mag_dh} shows the magnetar navigation results obtained by the Markov Chain Monte Carlo (MCMC) method, which are consistent with the true values. Note that Argument of Perigee ($\Omega$) and Mean Anomaly ($w$) are linearly negatively correlated due to the fact that the eccentricities ($e$) for Fermi and GECAM are close to 0, thus we cannot well limit these two elements; in fact only five elements are needed.

Table \ref{table2} lists the orbital elements obtained by this case of magnetar navigation. Although only 26 bursts from SGR 1935+2154 are used, and some of them are weak, the navigation results on both Fermi and GECAM orbits have accuracy of several hundred kilometers (see Table \ref{table2}). For spacecraft far away from the earth, e.g., in deep space, two well separated spacecrafts can detect most bursts without the blocking of the earth or other celestial bodies. We thus simulate 100 bursts (e.g. a burst `forest') for navigation and show that the errors of orbital elements can be about 20\% of the result for the 26 bursts. This indicates the validity and potential of this navigation method.

\section{Discussion and Conclusion}
In this work, we first performed pulsar navigation for the Fermi/GBM satellite. Although GBM is not designed for pulsar navigation and has relatively lower sensitivity, the orbital accuracy is still about 20 km using 16 days of Crab observations.
This provides a validation for future pulsar navigation using such small and low-cost detectors designed for other purposes. 

On the other hand, pulsar navigation has several difficulties, such as the need for long observation time and computational resources due to the relatively low flux of pulsars, as well as red noise and glitches (\citealp{lyne199323}; \citealp{scott2003characterization}; \citealp{hobbs2010analysis}; \citealp{lyne201545}). In addition, the SEPO method only considers the deviation of a single orbital parameter, however, deviation may exist among all the six orbital elements \citep{fang2021analysis}, which may lead to an overestimation of the orbital accuracy. Of course, a more comprehensive approach would be to free all orbital elements, such as those obtained by the MCMC method; however, this would consume more than orders of magnitude of computing resources.
Therefore, we also investigate the effects when another parameter is biased. As an example, let the deviation of $\Delta d$ be about 3 $\sigma$, and then calculate the result of $\Delta \omega$ as 0.228 (corresponding to 27.8 km), which is slightly larger than the value 0.216 (corresponding to 26.5 km) when $\Delta d$ is 0 (i.e. no deviation). This indicates that the deviations of other orbital elements do not have a very significant effect on this element, implying that SEPO method is not only convenient, but also effective.

Then we propose to use the repeated bursts for navigation based on  the time delays of the same bursts observed by multiple satellites, which is the inverse process of triangulation. In this work, we used the 26 bursts from SGR J1935+2154 observed by GBM and GECAM for navigation as a case of demonstration. Although only a few bursts are used, due to Earth occlusion and GECAM-B working for about only 11 hours per day during this time range, resulting in too few bursts being observed jointly by GBM and GECAM, the navigation still yielded orbital accuracy of several hundred kilometers.
Moreover, since letting all orbital elements free to fit at the same time, we can obtain the absolute and relative orbits of both satellites at the same time. It is worth noting that with this method, we can using any kinds of repeated bursts which can be detected by multi-satellite with good timing information.

The navigation method based on repeated bursts is more economic and convenient than pulsar navigation; the former uses only smaller and cheaper detectors, and the onboard calculations are also simple and fast without the need to correct the arrival times of events observed to the solar system barycenter, e.g., even in minutes on a personal computer, in comparison to traditional pulsar navigation 
requiring a powerful server. The main challenge for magnetar navigation is that the bursts from SGR J1935+2154 may exhibit aperiodic behavior \citep{xie2022revisit, zou2021periodicity}, i.e., in some periods there may not exist any burst. However, with future discoveries of more active magnetars (both X-ray and radio pulsars), this problem may be mitigated. According to a simulation in Section 3, the navigation accuracy can be reduced down to $\sim$ 100 km with 100 bursts. It is worth noting that the orbit accuracy obtained with this navigation method is robust, with the consideration of the deviations of all orbital elements (i.e. let all parameters be free).
Another issue worth noting is that some bursts may have spectral lags \citep{ukwatta2012lag,bernardini2014comparing,xiao2022robust}, that is, photons of different energies arrive at different times. This means that if the energy responses of the detectors on the two satellites differ significantly, it may introduce additional systematic errors in navigation. Fortunately, for the bursts from magnetars, the spectral lag is almost negligible \citep{10.1093/mnras/stad885}.
Finally, the method in fact does not need to know {\it a prior} the location of a bursting source; it just needs to include the location (i.e. Declination and Right Ascension) as additional parameters in the fitting. Therefore, this new navigation method has a good potential and can be combined with pulsar navigation for deep space exploration in the future.

\section*{Acknowledgments}
We thank the anonymous reviewer for a careful reading of our manuscript and suggestions. This work made use of the data from the {\it Fermi} and GECAM. This work is supported by the National Key R\&D Program of China (2022YFF0711404).
The authors thank supports from 
the Strategic Priority Research Program on Space Science, the Chinese Academy of Sciences (Grant No.
XDA15010100, XDA15360100, XDA15360102, XDA15360300,
XDA15052700) 
, the National Natural Science Foundation of China (Projects: 12061131007, Grant No. 12173038, Grant No. 12273008 and No. 62062025, 61662010), the Foundation of Education Bureau of Guizhou Province, China (Grant No. KY (2020) 003), Science and Technology Foundation of Guizhou Province (Key Program, No. [2019]1432, No. ZK[2022]304), the Scientific Research Project of the Guizhou Provincial Education (Nos. KY[2022]123, KY[2022]132, KY[2022]137), and the Major Science and Technology Program of Xinjiang Uygur Autonomous Region (No. 2022A03013-4). S. Xiao is grateful to W. Xiao, G. Q. Wang and J. H. Li for their useful comments.  
\bibliography{ref0}

\newpage
\appendix

\renewcommand\thefigure{\Alph{section}A\arabic{figure}}

\setcounter{figure}{0}  
\setcounter{table}{0} 

\renewcommand\thetable{\Alph{section}A\arabic{table}}

\begin{table*}[htbp]
\centering
\caption{\centering Time delays between GECAM-B and GBM obtained for a series of bursts from SGR J1935+2154.}\label{sgr_1935}
\begin{tabular}{cccc}

\hline
Number & Trigger time (UTC)      & True time delay (ms) & Calculated time delay (ms) \\
\hline
1      & 2021-09-10T01:04:33.500 & -27.5                & -26.3$\pm$0.9              \\
2      & 2021-09-10T05:35:55.500 & 7.4                  & 7.6$\pm$0.4                \\
3      & 2021-09-11T16:35:46.500 & -1.4                 & -1.0$\pm$1.9               \\
4      & 2021-09-11T16:39:21.000 & -3.1                 & -4.0$\pm$6.0               \\
5      & 2021-09-11T16:50:03.850 & -6.5                 & -6.2$\pm$0.3               \\
6      & 2021-09-11T17:01:09.800 & -6.7                 & -6.0$\pm$0.3               \\
7      & 2021-09-11T17:01:59.550 & -6.5                 & -8.2$\pm$12.0              \\
8      & 2021-09-11T17:04:29.800 & -5.9                 & -4.7$\pm$2.1               \\
9      & 2021-09-11T17:10:48.750 & -3.8                 & -4.2$\pm$0.7               \\
10     & 2021-09-11T18:54:36.050 & 1.4                  & 1.3$\pm$0.4                \\
11     & 2021-09-11T20:13:40.550 & -3.9                 & -3.8$\pm$0.4               \\
12     & 2021-09-11T20:22:59.050 & -0.1                 & -0.2$\pm$0.3               \\
13     & 2021-09-11T21:07:28.350 & 0.6                  & -9.0$\pm$5.8               \\
14     & 2021-09-11T22:51:41.600 & -4.7                 & -5.1$\pm$0.6               \\
15     & 2021-09-12T00:34:37.450 & -8.2                 & -7.4$\pm$0.3               \\
16     & 2021-09-12T00:45:49.400 & -7.2                 & -7.1$\pm$0.4               \\
17     & 2021-09-12T05:14:07.950 & -11.2                & -11.3$\pm$0.7              \\
18     & 2021-09-12T16:26:08.150 & -22.2                & -22.0$\pm$1.8              \\
19     & 2021-09-12T16:52:07.950 & 2.2                  & 2.7$\pm$2.8                \\
20     & 2021-09-13T00:27:25.200 & -28.0                & -28.2$\pm$0.3              \\
21     & 2021-09-13T19:51:33.350 & -12.7                & -13.0$\pm$0.7              \\
22     & 2021-09-14T11:10:36.250 & -10.9                & -10.6$\pm$0.2              \\
23     & 2021-09-14T14:15:42.900 & 4.1                  & 5.8$\pm$0.9                \\
24     & 2021-09-17T12:52:37.800 & 20.1                 & 23.4$\pm$2.4               \\
25     & 2021-09-17T13:58:25.100 & -25.3                & -24.4$\pm$3.8              \\
26     & 2021-09-18T22:58:52.150 & -4.5                 & -10.9$\pm$7.5 \\        \hline
\end{tabular}
\end{table*}

\begin{figure*}
    \centering
    \includegraphics[width=0.6\columnwidth]{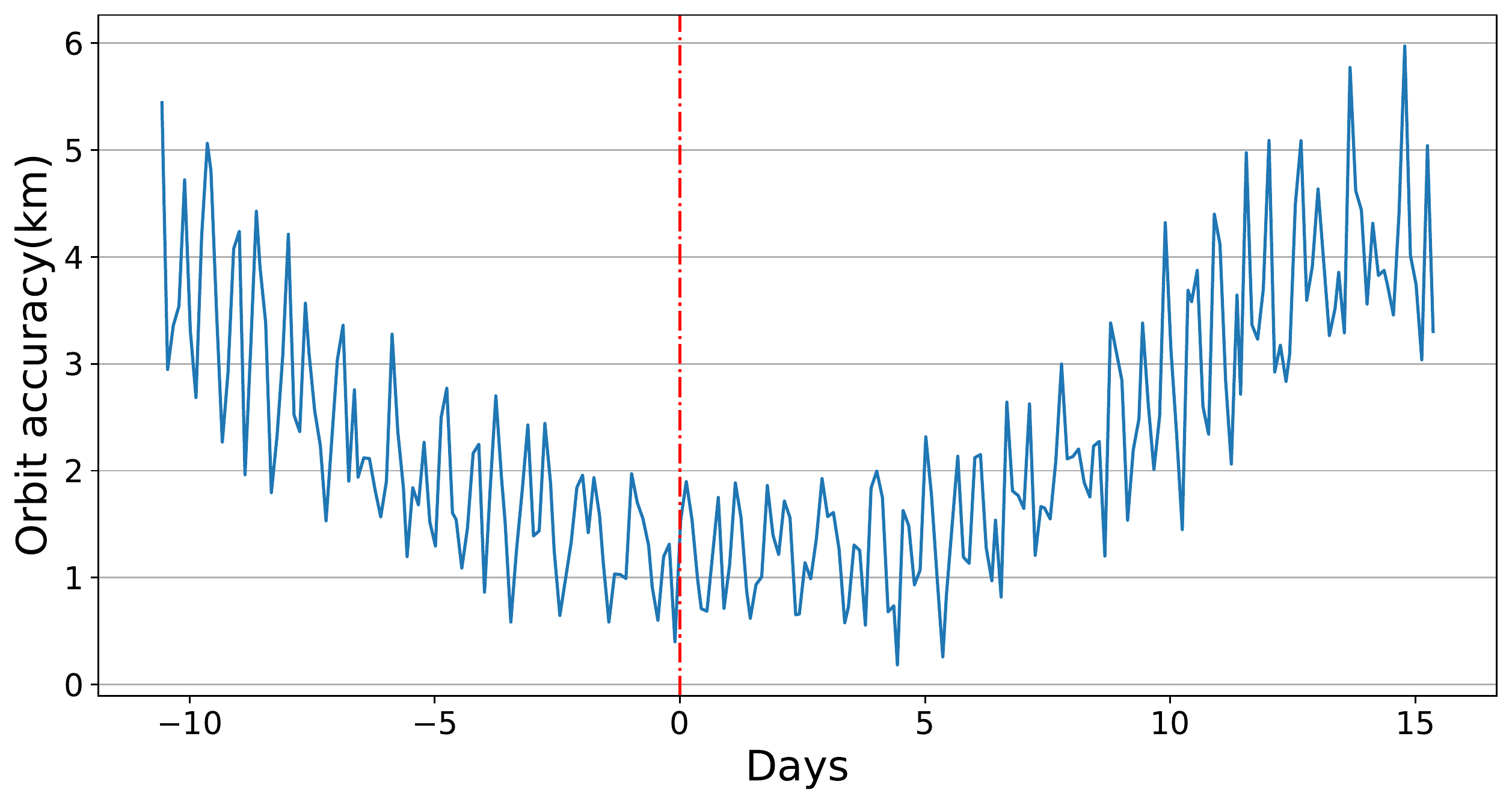}
    \caption{The difference between the predicted based on the dynamical model and true orbits at different times.}\label{A1}
\end{figure*}

\begin{figure*}
    \centering
    \includegraphics[width=\columnwidth]{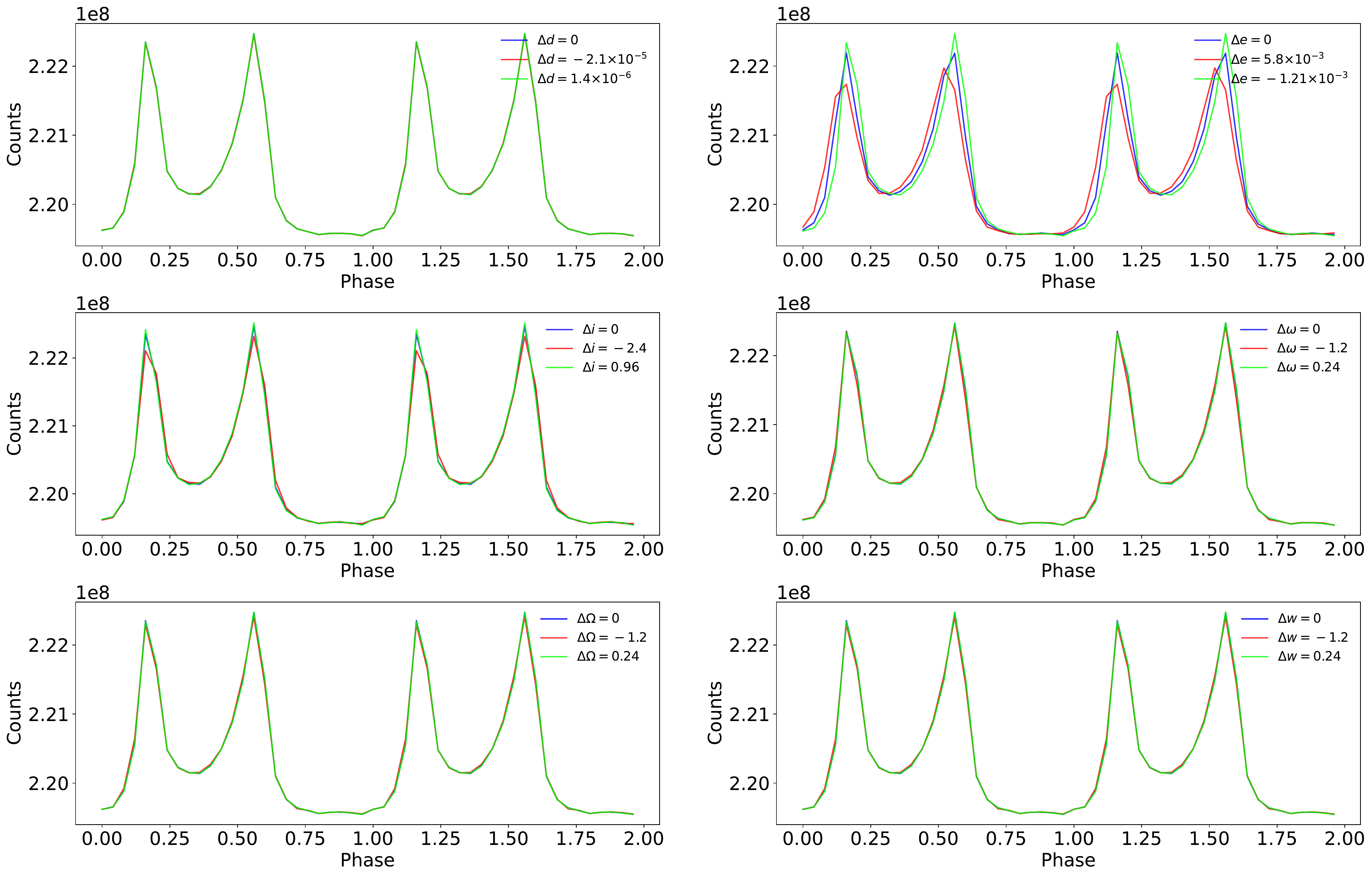}
    \caption{Pulse profiles obtained with different six orbital elements. Note due to statistical fluctuations, the maximum count may not be the highest when the deviation of element is 0 (i.e. true value).}\label{profiles_six}
\end{figure*}

\begin{table*}[htbp]
\centering
\caption{\centering Orbital elements of Fermi and GECAM obtained by magnetar navigation (using 26 bursts from SGR J1935+2154). The $e$, $\Omega$, $i$, $w$, $d$ and $\omega$ are the orbital elements of Fermi/GBM, respectively. The $e_1$, $\Omega_1$, $i_1$, $w_1$, $d_1$ and $\Omega_1$ are the orbital elements of GECAM-B, respectively.}\label{table2}
\begin{tabular}{c  c c c c}

\hline
Orbital Elements     &True values & Calculated values(1$\sigma$)  & Distance Error (km)
\\
\hline
$d${\rm (rad/s)}($10^{-3}$)     &62.983 &62.979 (-0.012, +0.013)   &(-70.933, +76.833)
\\
$e$                     &0.0012      &0.0035 (-0.0029, +0.0066)    &(-21.965, +51.868)             \\
$i$${\rm (^o)}$         &25.582        &23.244 (-9.710, +7.423)    &(-401.266, +293.841)            \\
$\omega$${\rm (^o)}$    &267.808       &272.208 (-15.181, +13.793) &(-1802.987, +1638.977)            \\
$\Omega$${\rm (^o)}$    &188.040       &185.665 (-16.452, +15.540) &(-1970.429, +1861.731)            \\
$w$${\rm (^o)}$         &171.995       &171.111 (-19.491, +18.033) &(-2346.187, +2172.769)            \\

$d_1${\rm (rad/s)}($10^{-3}$)     &62.119 &62.113 (-0.012, +0.012)  &(-71.471, +71.461) 
\\
$e_1$                    &0.0013        &0.0029 (-0.0021, +0.0044)  &(-22.245, +46.712)             \\
$i_1$${\rm (^o)}$        &28.999        &24.634 (-2.329, +4.500)    &(-164.749, +320.946) \\
$\Omega_1$${\rm (^o)}$   &58.185        &48.074 (-10.644, +12.874)  &(-1254.407, +1516.204) \\
$\Omega_1$${\rm (^o)}$   &251.199       &262.315 (-59.467, +62.507) &(-6900.146, +7215.313) \\
$w_1$${\rm (^o)}$        &108.721       &105.695 (-59.923, 78.885)  &(-6757.571, +8874.997) \\

\hline
\end{tabular} 

\footnotesize{$^*$ Note that Argument of Perigee ($\Omega$) and Mean Anomaly ($w$) are linearly negatively correlated due to the fact that Eccentricities ($e$) for Fermi and GECAM are close to 0, thus cannot well limit these two elements.}\\
\end{table*}

\end{CJK}
\end{document}